\documentclass{article}

   \usepackage[nonatbib, preprint]{neurips_2023}

\usepackage[utf8]{inputenc} % allow utf-8 input
\usepackage[T1]{fontenc}    % use 8-bit T1 fonts
\usepackage{hyperref}       % hyperlinks
\usepackage{url}            % simple URL typesetting
\usepackage{booktabs}       % professional-quality tables
\usepackage{amsfonts}       % blackboard math symbols
\usepackage{nicefrac}       % compact symbols for 1/2, etc.
\usepackage{microtype}      % microtypography
\usepackage{xcolor}         % colors
\usepackage{siunitx}
\usepackage{graphicx}
\usepackage{subcaption}

\title{Automated Model Training (AMT) GUI: An Opportunity for integrating AI in the Laboratory Experiment}

\author{
  Mohamed Bilal Shakeel
  \\Qatar Environment and Energy Research Institute and 
  \\College of Science and Engineering \\
  Hamad Bin Khalifa University\\
  Doha, Qatar \\
  \And
  Samir Brahim Belhaouari
  \\
  College of Science and Engineering\\
  Hamad Bin Khalifa University\\
  Doha, Qatar \\
  \And
  Fedwa El Mellouhi
  \\
  Qatar Environment and Energy Research Institute\\
  Hamad Bin Khalifa University\\
  Doha, Qatar \\
  \texttt{felmellouhi@hbku.edu.qa} \\
}

\begin{document}

\maketitle

\begin{abstract}
In the field of materials science, comprehending material properties is often hindered by the complexity of datasets originating from various sources. This study introduces the Automated Model Training (AMT) Graphical User Interface (GUI), specifically crafted for the use of researchers and scientists without a programming background to design the next set of experiments either in a chemistry lab or on the computer. The GUI integrates diverse machine learning models, such as XGBoost, Random Forest, Support Vector Regression, Linear Regression, Generalized Additive Model (GAM), and Stack Regressors, offering a robust toolkit for data analysis. It facilitates the exploration of complex relationships, non-linear patterns, and predictive accuracy optimization. To further enhance its utility, the GUI integrates the Particle Swarm Optimization (PSO) technique, allowing researchers to systematically explore vast parameter spaces and identify optimal experimental conditions. This synergy between machine learning and PSO empowers material scientists through a user-friendly platform for data-driven discovery. The AMT GUI bridges the gap between traditional experimentation and machine learning, enabling precise and efficient exploration of the materials research space. 
\end{abstract}

\section{Introduction}

In the realm of materials science, the inherent challenges associated with the prohibitively high costs and resource-intensive nature of laboratory experimentation have posed formidable obstacles in the pursuit of comprehensively exploring the multifaceted landscape of material properties. This terrain encompasses an array of variables, spanning diverse environmental conditions, complex compositional variations, intricate molar ratios, and the nuanced influence of supporting elements, among others. Recent advancements in the integration of machine learning methodologies have emerged as a transformative paradigm, effectively bridging the chasm between the constrained experimental capacity and the inexhaustible permutations of material configurations \cite{butler2018machine,schleder2019dft,chen2020machine}, hence enabling the systematic design of experiments.

Within this context, machine learning has demonstrated a pivotal role by enabling the emulation of experimental processes, offering precise prognostications for hitherto unexplored experimental scenarios \cite{butler2018machine} \cite{schmidt2019recent} \cite{chen2020machine}. Leveraging the computational prowess of machine learning, researchers and experimentalists can glean invaluable recommendations for optimal parameters and configurations. This synergy between materials science and machine learning, however, has accentuated a pronounced knowledge gap between practitioners with backgrounds in traditional experimental methodologies and the intricate nuances of machine learning techniques \cite{morgan2020opportunities,himanen2019data}.

In response to this exigency, we introduce the Automated Model Training (AMT) GUI, a bespoke tool meticulously tailored to the specific needs of researchers and scientists. The AMT GUI serves as an intuitive and user-friendly platform, empowering researchers to effortlessly train conventional machine learning models on their datasets. It further augments its utility by seamlessly integrating a computational optimization technique, Particle Swarm Optimization (PSO).

Moreover, the AMT GUI facilitates precise control over the exploration space for each  parameter, thereby empowering optimization algorithms to uncover extremal points within predefined ranges. This vital feature not only expedites the optimization process but also provides researchers with a robust mechanism to fine-tune their experiments and discern the most favorable conditions.

Researchers are currently developing interfaces to automate machine learning model training. Leading technology companies provide web-based graphical user interfaces (GUIs) that streamline this process, exemplified by Azure ML \cite{Lazzeri2019}. In the realm of open-source frameworks, platforms like KNIME \cite{KNIME} and Orange \cite{Orange} offer visual workflows designed to construct and train ML models. Additionally, specialized tools such as TPOT \cite{Team2018} focus on automating hyperparameter tuning, while Ludwig \cite{Ludwig} simplifies the training of deep learning models.

Among these solutions, Automated Model Training (AMT) stands out for its exceptional user-friendliness, specifically tailored for researchers and scientists without extensive programming backgrounds. Its interface is optimized for designing experiments, selecting models, and optimizing hyperparameters within the user's domain expertise.

What sets AMT apart is its strong commitment to user privacy. Unlike some cloud-based solutions that store experimental data on external servers, AMT ensures that all data remains securely stored on the user's local machine. This approach ensures maximum privacy and confidentiality throughout the entire training process.

This emphasis on usability, tailored scientific applications, and robust privacy protections positions AMT as a unique and indispensable tool for researchers aiming to accelerate their work while maintaining full control over their experiments and data.

\begin{figure}
  \centering
  \includegraphics[width=1\textwidth]{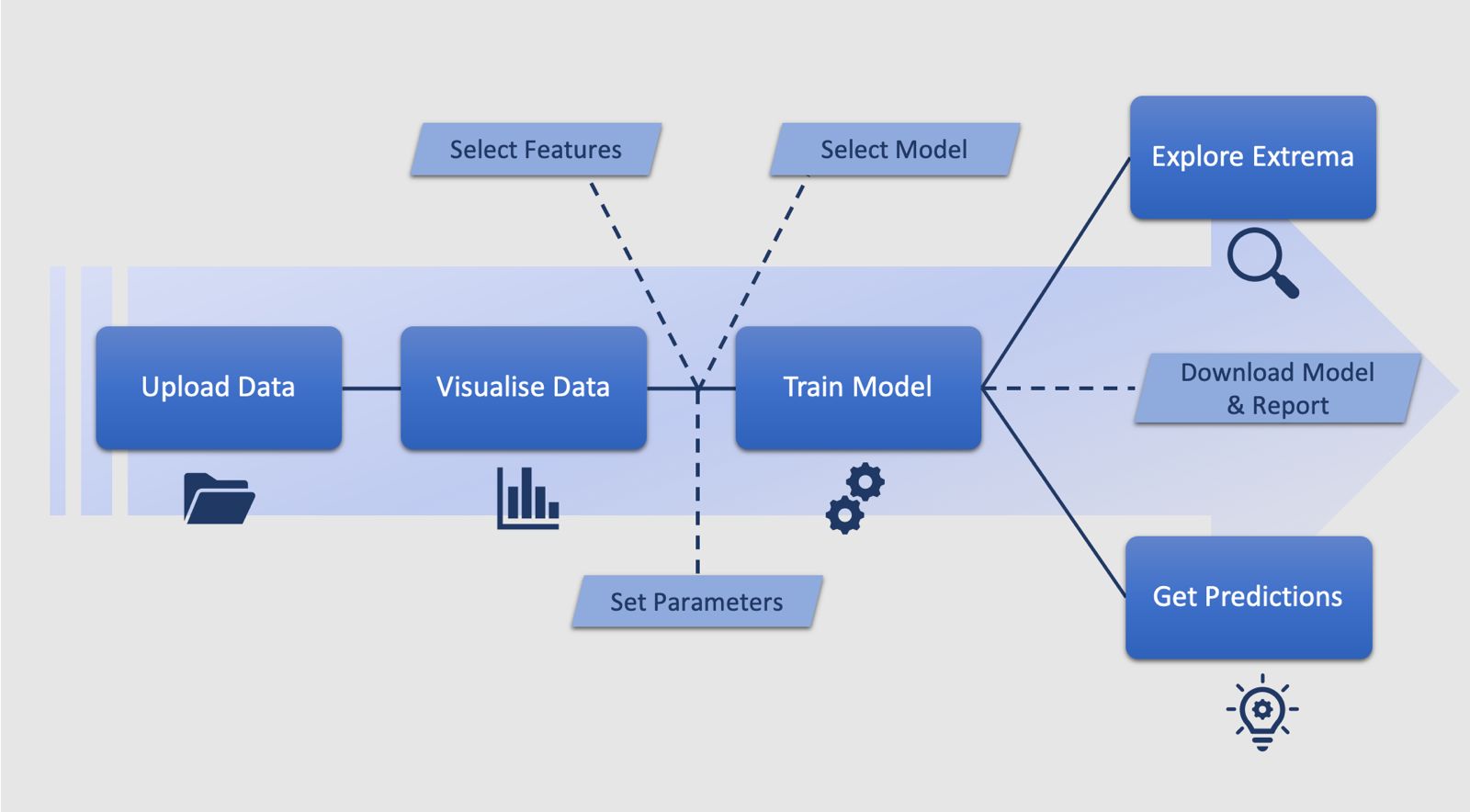}
  \caption{Flow Diagram Illustrating the Operational Steps of the AMT GUI}
  \label{fig:mainfigure}
\end{figure}

\section{AMT Design and Functionality}
\subsection{GUI Overview}

The development of the Automated Model Training (AMT) GUI incorporates ReactJS as the frontend framework, strategically selected for its prowess in rendering and updating mechanisms, bolstered by the utilization of a Virtual DOM \cite{rawat2020reactjs}. This architectural choice enables the GUI to selectively update only the relevant components of the user interface that have undergone changes, thereby bestowing exceptional performance capabilities. On the backend, Python serves as the integral language for executing all machine learning and optimization algorithm processes, thus forming the backbone of the system's computational functions. This symbiotic fusion of frontend and backend technologies ensures a robust, high-performance tool for material scientists and researchers.

Figure \ref{fig:mainfigure} presents a comprehensive flow diagram encapsulating the primary operational steps within the AMT GUI (Graphical User Interface). Initially, the process commences with the user uploading a datafile in a prescribed format, thereby initializing the analytical journey. Subsequently, the interface furnishes the user with a suite of visualization tools, facilitating an in-depth exploration of the dataset through diverse graphical representations.

Following this exploratory phase, users are empowered to engage in feature manipulation, affording them the capability to selectively include or exclude specific features based on the insights gleaned from the visualizations. This critical step serves as a precursor to the model selection phase, wherein users are presented with a repertoire of machine learning models to choose from. Moreover, for users inclined towards fine-tuning model performance, the GUI extends the functionality to customize model parameters, ensuring a tailored approach to model training.

Upon satisfaction with the training results, users are presented with an array of post-training actions. These encompass predictive analytics, wherein the trained model is employed to make predictions, thereby facilitating informed decision-making. Furthermore, users are presented with the opportunity to delve deeper into the model's performance within a restricted search space. This capability facilitates the exploration of optimized values within specified parameters, thereby enabling a more nuanced understanding conducive to designing laboratory experiments.

Concluding the workflow, the GUI offers users the flexibility to download comprehensive reports encapsulating model performance metrics and insights. Additionally, users can elect to export the trained model for future utilization or integration into alternative applications, thus ensuring the scalability and interoperability of the analytical framework.

\subsection{Dataset Selection for GUI Demonstration}
In the remainder of the paper, we offer an in-depth examination of the utilization and efficacy of the AMT GUI in conjunction with an exemplary dataset pertaining to Concrete Compressive Strength \cite{misc_concrete_compressive_strength_165}. The dataset encompasses a total of 1030 instances, featuring 8 quantitative input variables and 1 quantitative output variable. The input variables include Cement (\si{kg/m^3}), Fly ash (\si{kg/m^3}), Blast furnace slag (\si{kg/m^3}), Water (\si{kg/m^3}), Superplasticizer (\si{kg/m^3}), Coarse aggregate (\si{kg/m^3}), Fine aggregate (\si{kg/m^3}), and Age of testing (\si{days}). These attributes collectively contribute to the prediction of concrete compressive strength, a critical parameter in assessing the performance and durability of concrete structures.The properties of these components, including their minimum, maximum, and average values, are summarized in Table \ref{tab:component_properties}. This table provides a comprehensive overview of the ranges and typical values for each component within the dataset, aiding in the understanding of their influence on concrete compressive strength prediction.

\begin{table}[htbp]
    \centering
    \caption{Properties of Components in the Dataset}
    \begin{tabular}{lccc}
        \toprule
        Component & Minimum (kg/m$^3$) & Maximum (kg/m$^3$) & Average (kg/m$^3$) \\
        \midrule
        Cement & 102 & 540 & 281.16 \\
        Blast furnace slag & 0 & 359.4 & 54.18 \\
        Fly ash & 0 & 200.1 & 73.89 \\
        Water & 121.8 & 247 & 181.56 \\
        Superplasticizer & 0 & 32.2 & 6.20 \\
        Coarse aggregate & 801 & 1145 & 972.91 \\
        Fine aggregate & 594 & 992.6 & 773.58 \\
        Age (day) & 1 & 365 & 45.66 \\
        Concrete compressive strength (MPa) & 2.33 & 82.6 & 35.81 \\
        \bottomrule
    \end{tabular}
    \label{tab:component_properties}
\end{table}

\subsection{Data Upload and Insights}
Uploading experimental data for processing in AMT adheres to a user-friendly predefined template, readily available for download within the GUI. Once the data file is correctly formatted and uploaded, the full spectrum of features becomes accessible. Any format-related errors are promptly flagged and communicated in the status notification.

\begin{figure}
  \centering
  \includegraphics[width=1\textwidth]{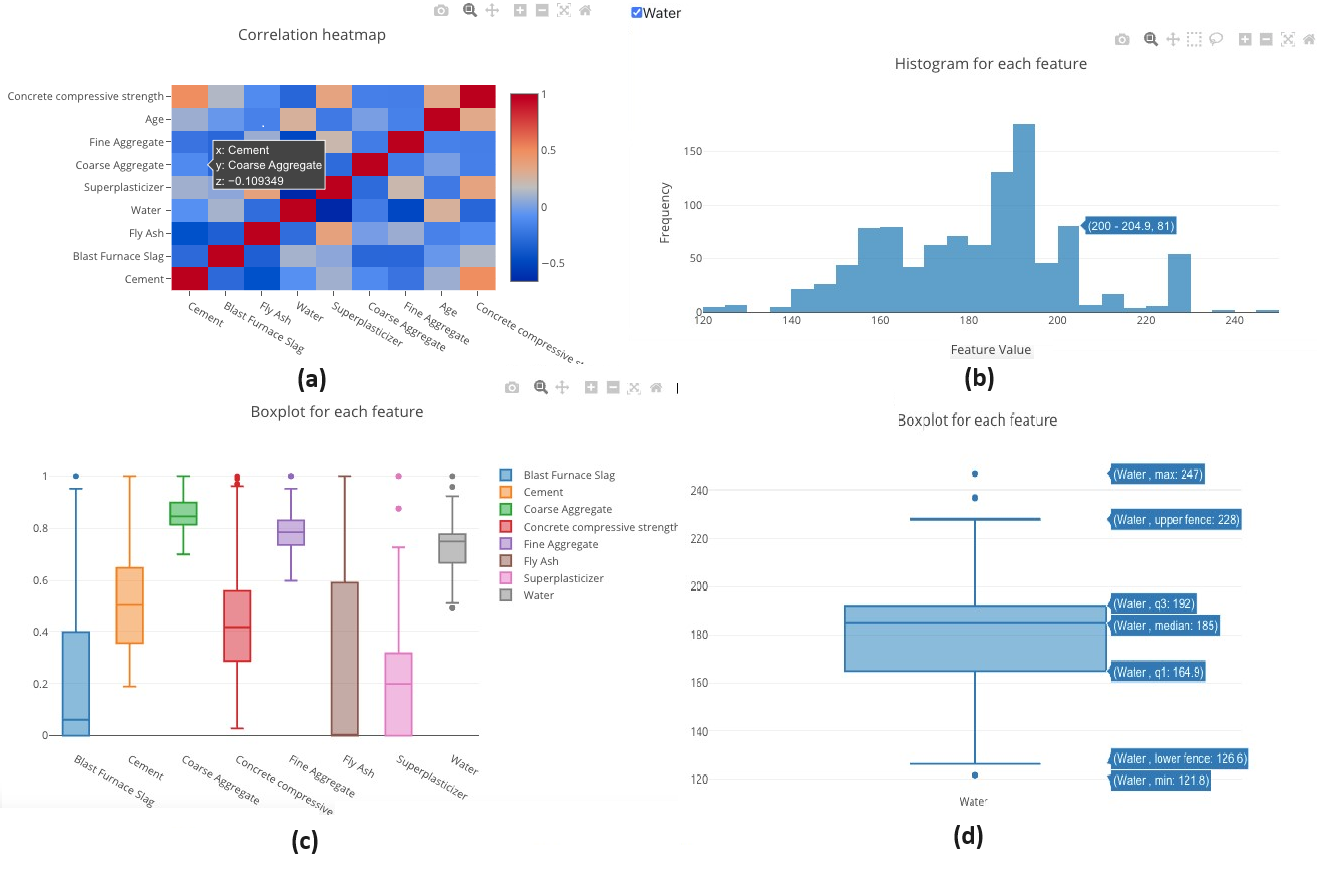}
  \caption{GUI Empowered Visualizations: Unveiling Data Insights through Correlation Heatmaps (a), Histograms (b), and Boxplots (c) and (d). Each visualization tool offers a distinct lens into the exemplary dataset, facilitating comprehensive inspection and analysis of individual inputs.}
  \label{fig:visualization}
\end{figure}

\begin{figure}
  \centering
  \begin{subfigure}{0.49\textwidth}
    \includegraphics[width=\textwidth]{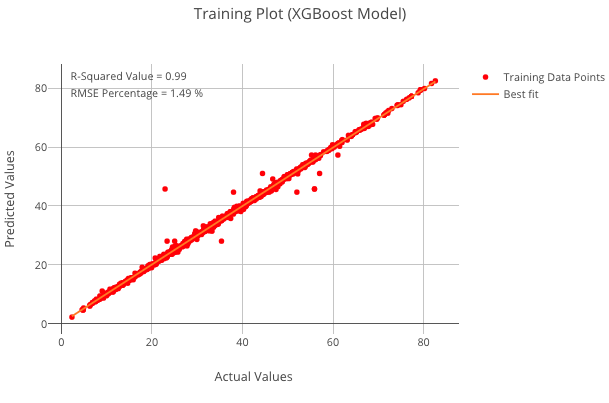}
    \subcaption{}
    \label{fig:subfigurea2}
  \end{subfigure}
  \hfill
  \begin{subfigure}{0.49\textwidth}
    \includegraphics[width=\textwidth]{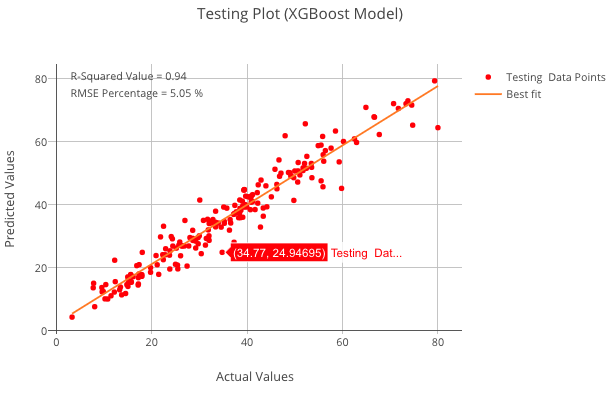}
    \subcaption{}
    \label{fig:subfigureb2}
  \end{subfigure}
\caption{Evaluation of the XGBoost-trained model's performance on the exemplary dataset, divided into training (a) and testing (b) datasets with 90-10 split ratio, showcasing high R-squared values and low RMSE (Root Mean Squared Error), indicative of strong predictive accuracy and minimal model error.}

  \label{fig:mainfigure2}
\end{figure}

To furnish valuable insights from the uploaded experimental data, the GUI offers a suite
of interactive visualization tools, including histograms, correlation heatmaps and boxplots. Figure \ref{fig:mainfigure2} showcases these visualizations for the exemplary dataset. Histograms provide an overarching perspective on the distribution of experiment data. If skewness is detected in either direction, it signals potential performance discrepancies in the associated areas. Correlation heatmaps, on the other hand, depict the relationships among input variables and the output variable, plays a pivotal role in optimizing the modeling process. It unveils the strength and direction of these relationships, facilitating the identification of influential predictors for model construction. Beyond this, it assists in the informed selection of the most suitable machine learning algorithm. For instance, if robust linear relationships emerge between inputs and outputs, Linear Regression becomes an evident choice. Conversely, when dealing with non-linear relationships, decision tree-based models like Random Forests or Gradient Boosting Trees offer enhanced predictive capabilities. In cases of high multicollinearity, regularization techniques such as Ridge or Lasso Regression come to the forefront, bolstering model stability. Moreover, for datasets exhibiting a medley of linear and non-linear relationships, ensemble methods like XGBoost, which amalgamate multiple models, prove adept at accommodating the diverse spectrum of relationship types. This multifaceted approach ensures that the modeling strategy is precisely aligned with the underlying data structure, promoting superior predictive performance. Boxplots, meanwhile, provide a concise summary of the distribution of data, offering insights into central tendency, variability, and potential outliers within each variable. By visually representing these statistical measures, boxplots enable users to quickly discern patterns and variations in the data, aiding in the identification of anomalous observations and guiding the selection of appropriate modeling techniques.

\subsection{Machine Learning Models}
The selection and integration of machine learning models within AMT GUI tailored for material science researchers hold paramount significance in advancing the field's data analysis capabilities. The rationale behind including models such as XGBoost \cite{ester2022xgboost}, Random Forest \cite{liu2012new} Support Vector Regression (SVR) \cite{zhang2020support}, Linear Regression \cite{su2012linear}, Generalized Additive Models (GAM) \cite{hastie1992generalized}, and Stack Regressors \cite{pavlyshenko2018using} is deeply rooted in the multifaceted nature of materials data and the unique challenges posed by this domain.

\begin{itemize}
  \item XGBoost and Random Forest are prized for their ability to handle non-linear relationships and intricate interactions, which are prevalent in materials datasets. These ensemble methods excel in capturing complex patterns and feature interactions, thereby offering researchers a powerful tool for predictive modeling and feature importance assessment.

  \item Support Vector Regression (SVR) is specially tailored for high-dimensional datasets, a prevalent trait in materials research. Its aptitude for managing intricate data structures and discerning meaningful patterns renders it invaluable for extracting valuable insights from multifaceted materials datasets.

  \item Linear Regression, while seemingly straightforward, plays a vital role in elucidating simple linear dependencies within materials datasets. Its interpretability makes it a valuable asset for researchers seeking to understand fundamental relationships between variables.

  \item Generalized Additive Models (GAMs) offer a versatile approach, allowing for the modeling of both linear and non-linear relationships. In materials science, where complex and non-linear phenomena often prevail, GAMs provide a flexible framework for capturing intricate patterns.

  \item Stack Regressors bring an additional layer of sophistication by combining the predictive power of multiple models. This ensemble approach enhances predictive accuracy and robustness, providing researchers with a valuable tool for improving the reliability of their predictions.
\end{itemize}

The GUI offers user-friendly access to essential parameters for fine-tuning each machine learning model. For users unfamiliar with these parameters, default settings are readily available, simplifying the training process. Post-training, graphical representations of model performance on the training and testing data are presented, as exemplified in Figure \ref{fig:mainfigure2}. Evaluation metrics, specifically the R-squared (R$^2$) and Root Mean Square Error (RMSE), provide valuable insights into model performance. A higher R-squared value indicates better model training, while the RMSE metric operates inversely, with a lower value signifying superior model performance.

In Figure \ref{fig:mainfigure2}, the training plot reveals an exceptional R-squared value of 0.99, coupled with an RMSE percentage of 1.49\%. This observation prompts consideration of two potential scenarios: either the XGBoost model exhibits an exceptional fit to the training data, or it is potentially suffering from overfitting. To discern between these possibilities, the model undergoes testing on unseen data. Impressively, the evaluation on the testing data yields a notable R-squared value of 0.94 and an RMSE value of 5.05\%. These results underscore the robustness of the model's fit to the uploaded data, providing validation of its predictive capabilities.

\begin{figure}
  \begin{minipage}[c]{0.27\textwidth}
    \includegraphics[width=\textwidth, height=9cm]{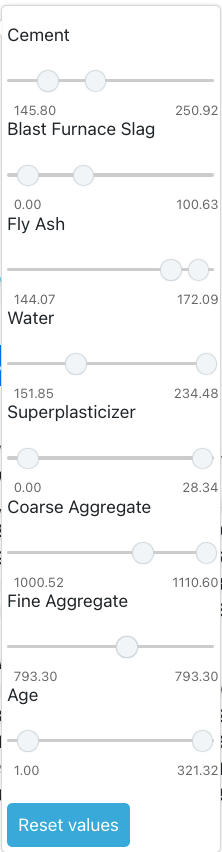} 
    \subcaption{}
    \label{fig:subfigurea3}
  \end{minipage}
  \hfill
  \begin{minipage}[c]{0.85\textwidth}
    \begin{minipage}[c]{\textwidth}
      \includegraphics[width=\textwidth]{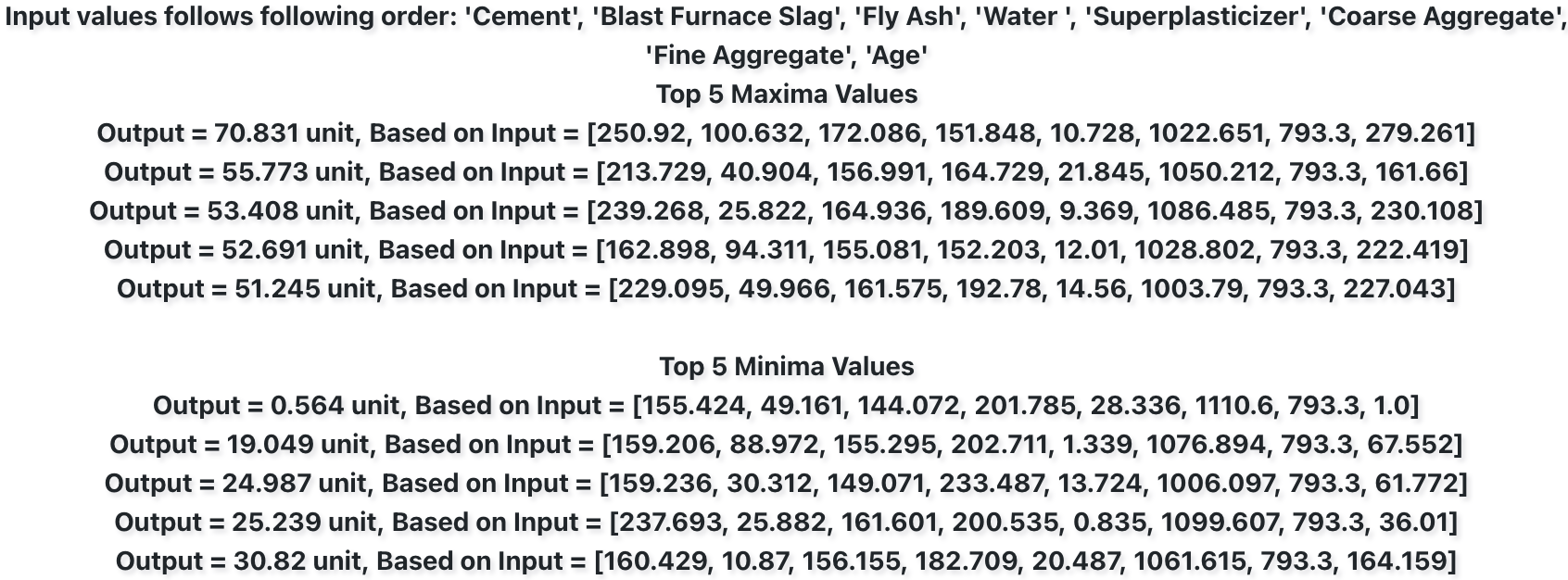}
      \subcaption{}
      \label{fig:subfigureb3}
    \end{minipage}
    \vfill

  \end{minipage}
  \caption{(a) Illustrates the flexibility of customizing input parameter boundaries (b) PSO solution across the entire exploration space }
  \label{fig:personalizedpso}
\end{figure}

\subsection{Optimization Algorithm}

In addition to the diverse ensemble of machine learning models, the AMT GUI also incorporates the Particle Swarm Optimization (PSO) \cite{wang2018particle} technique, further enriching its capabilities. PSO is a powerful optimization algorithm that operates based on the collective behavior of particles in a multidimensional search space. Numerous studies have underscored the versatile applications of Particle Swarm Optimization (PSO) across various domains, spanning from engineering and environmental optimization challenges to the fields of machine learning and bioinformatics 
\cite{shami2022particle,gad2022particle,pervaiz2021systematic,wang2021particle,wang2022reinforcement,jahandideh2020application}.
Its inclusion in the AMT holds substantial significance for material science researchers. PSO facilitates the derivation of optimal solutions by systematically exploring the vast parameter space encompassing environmental conditions, compositional variations, and material properties. This capability proves invaluable to researchers seeking to identify the most favorable conditions for their experiments. Figure \ref{fig:subfigureb3}  displays a screenshot captured from the AMT GUI, showcasing the top 5 maxima and top 5 minima solutions generated by the Particle Swarm Optimization (PSO) algorithm. These results are obtained using a trained XGBoost model on the Concrete Compressive Strength  dataset. In AMT, the inputs are listed in sequential order from left to right. In this case, it encompasses all the inputs mentioned in Table \ref{tab:component_properties}. 
Upon analyzing Figure \ref{fig:subfigureb3} within the context of our dataset, it is evident that the Particle Swarm Optimization (PSO) algorithm attains a maximum output of 70.831 units, corresponding to a specific set of input values. A detailed examination of these suggested inputs reveals their practical plausibility. For instance, the cement-to-water ratio is calculated to be 1.652 (250.92 (\si{kg/m^3}) cement per 151.848 (\si{kg/m^3}) of water), a proportion conducive to the formation of strong concrete. Moreover, the moderate inclusion of superplasticizer corroborates the model's validity. Superplasticizers are known to enhance concrete workability without augmenting the water content, thereby maintaining a low water-to-cement ratio, which can potentially increase compressive strength \cite{smith2020effect, alhassan2024production}. These additives function by dispersing fine particles within the mix, thereby enhancing the material's workability for placement, handling, and consolidation \cite{smith2020effect}. However, the relationship between superplasticizer dosage and concrete compressive strength is intricate, influenced by numerous factors. While superplasticizers can elevate compressive strength by improving compaction efficiency to produce denser concrete \cite{alsadey2022effect}, excessive dosages may conversely reduce compressive strength \cite{gagne1996effect}. Empirical evidence indicates that dosages exceeding 1.1 percent can diminish compressive strength by 15 to 50 percent \cite{gagne1996effect}.
Conversely, the algorithm also identifies the input configuration that results in the minimum output, recorded at 0.564 units. This outcome is logical, given the extremely low cement-to-water ratio of 0.77, which would inevitably yield very weak concrete.

Notably, it is essential to underscore that PSO operates on real-number values. In practice, these decimal values can be conveniently approximated by selecting the nearest rounded numerical equivalents wherever required. Furthermore, PSO offers users the flexibility to individually restrict the exploration space for each parameter, facilitating optimization within personalized boundaries, as illustrated in Figure \ref{fig:subfigurea3}. This capability is particularly valuable in scenarios where resource constraints hinder researchers from conducting laboratory experiments based on PSO’s recommendations, given the algorithm’s propensity for suggesting extreme parameter values in the optimization process. By harnessing PSO within the GUI, material scientists can efficiently fine-tune their experiments, expedite the discovery of optimal material configurations, and navigate the intricate landscape of materials research with greater precision and efficiency. The synergy between machine learning models and PSO offers an integrated platform that empowers researchers to make informed decisions, optimize experimental conditions, and extract valuable insights from their datasets.

\section{User Testing and Feedback on the AMT: Insights and Recommendations}

Following the development of the AMT GUI, a diverse group of researchers spanning various backgrounds, ranging from computer engineering to chemical engineering, and possessing qualifications ranging from bachelor's degrees to scientists, engaged in rigorous testing and feedback sessions with the tool. Two datasets were made available for AMT GUI testing and feedback purposes: a smaller dataset containing 9 samples and 4 features, and a larger dataset containing 12,708 samples and 11 features.

  Despite a lack of programming experience, all participants successfully employed the AMT GUI to train machine learning models on these datasets. The ensuing feedback can be summarized as follows:

\begin{itemize}

  \item User-Friendliness and Value: The AMT GUI was consistently lauded for its exceptional user-friendliness and value as a research tool. Users expressed a genuine preference for incorporating the AMT GUI into their future research endeavors.

  \item Data Input Flexibility: Some users suggested an enhancement in the way prediction values could be entered after model training. While the tool currently utilized range sliders, users expressed a desire for the option to input values via keyboard input as well.

  \item Historical Reporting: Beyond the generation of model training reports, feedback also encompassed a request for the provision of historical reports detailing the outcomes of fine-tuning exercises. This feature was deemed valuable for research purposes.

  \item Automated Parameter Tuning: The final feedback highlighted the potential for further improvement. Users recommended that, in addition to providing basic parameters for user-driven fine-tuning, the AMT GUI could enhance its functionality by automating the process of experimenting with different parameter settings autonomously.
\end{itemize}

These insightful suggestions and feedback from our diverse pool of testers not only underscore the user-friendliness and value of the AMT GUI but also shed light on avenues for future enhancements that can bolster its utility in the research community.

\section{Acknowledgement}

This work is supported by the Hamad Bin Khalifa University Vice President of Research Qatar Thematic grant cycle 1. project number VP-TG06. National Research Fund (QNRF) through the National Priorities Research Program (NPRP) under project number NPRP12S-0209-190063. 

\small
%%%%%%%%%%%%%%%%%%%%%%%%%%%%%%%%%%%%%%%%%%%%%%%%%%%%%%%%%%%%
%\bibliography{references}

\end{document}